\documentclass[prb,aps,twocolumn,amstex,preprintnumbers,amsmath,amssymb,array,tabularx]{revtex4}
\usepackage{graphicx}
\usepackage{dcolumn}
\usepackage{bm}
\usepackage{lscape} 
\usepackage{amsmath}
\usepackage{amsxtra}
\usepackage{tabularx}

\DeclareGraphicsExtensions{.jpg,.pdf,.eps,.gif}

\begin{document}

\title{Half-metallic graphene nanodots}

\author{$\mbox{Oded Hod}^1$, $\mbox{Ver\'onica Barone}^2$, and
  $\mbox{Gustavo E. Scuseria}^1$}

\affiliation{$^1$Department of Chemistry, Rice University, Houston,
  Texas 77005-1892;\\ $^2$Department of Physics, Central Michigan
  University, Mt. Pleasant, MI 48859}

\date{\today}

\begin{abstract}
  A comprehensive {\it first-principles} theoretical study of the
electronic properties and half-metallic nature of finite rectangular
graphene nanoribbons is presented.  We identify the bisanthrene isomer
of the C$_{28}$H$_{14}$ molecule to be the smallest polycyclic
aromatic hydrocarbon to present a spin polarized ground state.  Even
at this quantum dot level, the spins are predicted to be aligned
antiferromagnetically at the two zigzag edges of the molecule.  As a
rule of thumb, we find that zigzag edges that are at least three
consecutive units long, will present spin polarization if the width of
the system is $1$~nm or wider.  Room temperature detectability of the
magnetic ordering is predicted for molecules with zigzag edges $1$~nm
and longer.  For the longer systems studied, spin wave structures
appear in some high spin multiplicity states.  Energy gap oscillations
with the length of the zigzag edge are observed.  The amplitude of
these oscillations is found to be smaller than that predicted for
infinite ribbons.  The half-metallic nature of the ribbons under an
external in-plane electric field is found to be preserved even for
finite and extremely short systems.

\end{abstract}

\maketitle
 


Since their recent successful fabrication,~\cite{Geim2004} graphene
nanoribbons (GNRs) have been the focus of extensive experimental and
theoretical efforts.  GNRs have the same unique hexagonal carbon
lattice as carbon nanotubes (CNTs), confined to a
quasi-one-dimensional structure. Hence, they share a variety of
interesting physical characteristics.  Experimental evidence of
ballistic electronic transport, large phase coherence lengths, and
current density sustainability,~\cite{deHeer2006} accompanied by
theoretical predictions and experimental verification of interesting
magnetic properties~\cite{Fujita1996, Wakabayashi1998,
  Wakabayashi1999, Kusakabe2003, Yamashiro2003, Lee2005, Son2006},
quasi-relativistic behavior,~\cite{Zhang2005, Peres2006, Peres2006-2,
  Novoselov2007} and bandgap engineering capabilities~\cite{Ezawa2006,
  Barone2006, Son2006-2, Han2007} mark GNRs as potential building
blocks in future nanoelectronic devices.  Furthermore, due to their
planar geometry, standard lithographic techniques may be used for the
flexible design of a variety of experimental devices~\cite{Geim2004,
  Zhang2005, deHeer2006, Novoselov2007, Han2007} in a controllable and
reproducible manner.

Despite the aforementioned similarities, there is a distinct
difference between CNTs and GNRs.  Unlike the tubular shaped
nanotubes, GNRs, which are long and narrow strips cut out of a two
dimensional graphene sheet, present long and reactive edges prone to
localization of electronic states.  The importance of these edge
states was demonstrated for the case of armchair CNTs which present a
metallic and non-magnetic~\cite{saito_book, dress-book} character in
their tubular form, but when unrolled into the corresponding zigzag
GNRs, they were predicted to become semiconducting~\cite{Nakada1998,
  Son2006} with a spin polarized~\cite{Fujita1996, Wakabayashi1998,
  Wakabayashi1999, Kusakabe2003, Yamashiro2003, Lee2005, Son2006}
ground state.  This ground state is characterized by opposite spin
orientations of localized electronic states at the two edges of the
GNR, which couple through the graphene backbone via an
antiferromagnetic (AF) arrangement of spins on adjacent atomic sites.

In a recent study, Son {\em et al.}~\cite{Son2006} have shown that
upon the application of an electric field, an opposite local gating
effect of the spin states on the two edges of the ribbon may occur.
The in-plane field (perpendicular to the periodic axis of the ribbon)
drives the system into a half-metallic state where one spin flavor
exhibits metallic behavior, while the opposite experiences an
increase in the energy gap.  Chemical doping of the ribbon edges was
shown to enhance this effect resulting in an efficient and robust spin
filter device.~\cite{Hod2007}

Following these studies on the half-metallic nature of {\em
  one-dimensional} zigzag graphene nanoribbons, several papers on the
electronic properties,~\cite{Silvestrov2007} magnetic
properties~\cite{Shemella2007, Ezawa2007, Rossier2007, Jiang2007} and
the half-metallic nature~\cite{Rudberg2007, Kan2007} of quasi-{\em
  zero-dimensional} graphene-based structures have appeared.  These
concluded that the spin polarized character of the zigzag graphene
edges persists also in nanometer scale islands of graphene, and that
the half-metallic nature of the structures studied by Son {\em et
  al.}~\cite{Son2006} may be an artifact of the level of theory they
applied.~\cite{Rudberg2007}

\input{epsf}
\begin{figure}[h]
\begin{center}
\epsfxsize=8.5cm \epsffile{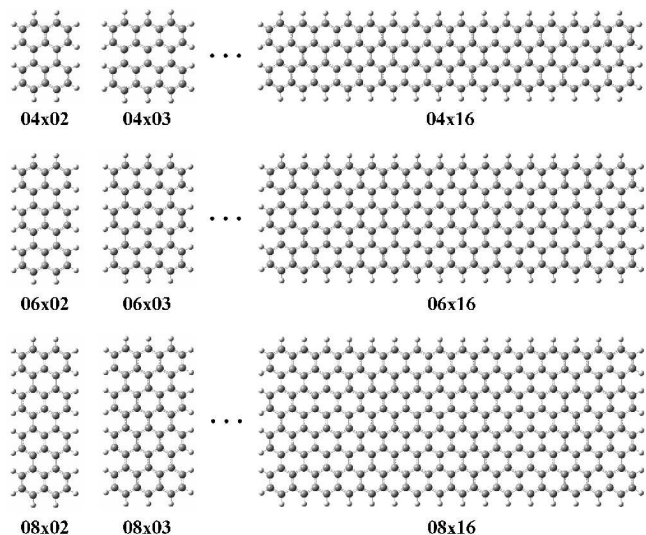}
\end{center}
\caption{The three sets of finite GNRs studied. The notation
  corresponds to the number of hydrogen atom passivating the edges. A
  $N\times M$ ribbon has $N$ hydrogens on the armchair edge and $M$
  hydrogens on the zigzag edge.}
\label{Fig: Finite Ribbons}
\end{figure}

It is the purpose of our paper to present a comprehensive and
systematic analysis of the electronic properties of finite-sized
graphene nanoribbons.  We present a general rule of thumb for the
existence of a spin-polarized ground state in aromatic carbon based
materials and identify the molecules C$_{28}$H$_{14}$
(Phenanthro[1,10,9,8-opqra]perylene) and C$_{36}$H$_{16}$
(Tetrabenzo[bc,ef,kl,no]coronene) as the smallest hydrocarbon
structures to possess a magnetic ground state.  The stability of the
magnetic ordering, the energy gap dependence on the dimensions of the
system, and the effect of an externally applied electric field are
also studied.  Unlike previous studies,~\cite{Rudberg2007} we show
that the half-metallic nature of the ribbons (whether periodic or
finite) is robust and insensitive to the level of theory used.

To this end we study a large set of finite rectangular nanoribbons of
different widths and lengths using three levels of density functional
approximations.  We notate the ribbons according to the number of
hydrogen atoms passivating the edges, such that a $N\times M$ finite
ribbon has $N$ hydrogen atoms on its armchair edge and $M$ atoms on
its zigzag edge (see Fig.~\ref{Fig: Finite Ribbons}).  The ribbons we
consider constitute three sub-sets corresponding to three ribbon
widths: $4\times M$, $6\times M$, and $8\times M$, where $M=2,...,16$.
All the calculations presented in this work were carried out using the
development version of the {\it Gaussian} suite of
programs.~\cite{gdv_short} Spin polarized ground state calculations
were performed using the local density approximation~\cite{Keywords}
(LDA), the semi-local gradient corrected functional of Perdew, Burke
and Ernzerhof~\cite{pbe_1,pbe_2,Keywords} (PBE), and the screened
exchange hybrid density functional due to Heyd, Scuseria and Ernzerhof
(HSE06),~\cite{hse, hse-errata, Izmaylov2006, Keywords} which has been
tested in a wide variety of materials and has been shown to accurately
reproduce experimental band gaps~\cite{hse-bulk,hse-bulk2} and first
and second optical excitation energies in metallic and semiconducting
single walled CNTs.~\cite{chirals,metallic} The inclusion of
short-range exact-exchange in the HSE06 functional makes it suitable
to treat electronic localization effects~\cite{Kudin2002, Prodan2005,
Prodan2006, Hay2006, Kasinathan2006} which are known to be important
in this type of materials.~\cite{Kobayashi1993, Fujita1996,
Nakada1996, Wakabayashi1998, Nakada1998, Wakabayashi1999,
Miyamoto1999, Kawai2000, Okada2001, Kusakabe2003, Yamashiro2003,
Niimi2005, Kobayashi2005, Lee2005, Son2006, Son2006-2, Niimi2006,
Kobayashi2006,Hod2007} This is further supported by the good agreement
which was recently obtained between predicted
bandgaps~\cite{Barone2006} of narrow nanoribbons and measured
values.~\cite{Han2007} To obtain a reliable ordering of the different
magnetization states we find it important to relax the geometry of the
finite GNRs for each spin polarization.  Therefore, unless otherwise
stated, all reported electronic properties are given for fully
optimized structures for each approximate functional using the
polarized 6-31G** Gaussian basis set.~\cite{Hariharan1973} It should
be noted that since our calculations are performed within a single
determinantal framework, we can determine only the total spin vector
projection along a given axis $m_s$ and not the total spin.

\begin{figure}[h]
\begin{center}
\epsfxsize=8.5cm \epsffile{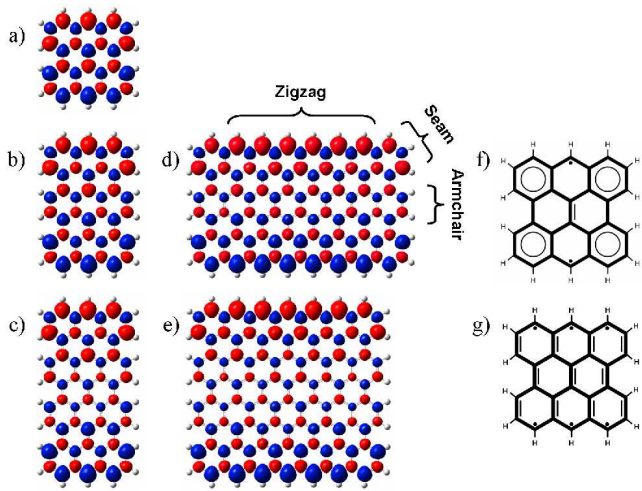}
\end{center}
\caption{Isosurface spin densities of the antiferromagnetic ground
  state of the 04x03 (panel (a)), 06x03 (panel (b)), 08x03 (panel
  (c)), 06x08 (panel (d)), and 08x07 (panel (e)) GNRs as obtained
  using the HSE06 functional and the 6-31G** basis set.  The sorting
  of the atoms into zigzag, armchair and seam regions is indicated in
  panel (d).  Panels (f) and (g) represent the diradical and
  hexaradical Clar structures of bisanthrene}
\label{Fig: Spin Density}
\end{figure}

Given that most organic molecules are known to be diamagnetic, while
periodic~\cite{Fujita1996, Wakabayashi1998, Wakabayashi1999,
  Kusakabe2003, Yamashiro2003, Lee2005, Son2006, Hod2007} and finite
graphene clusters~\cite{Rudberg2007, Shemella2007, Ezawa2007,
  Rossier2007} are predicted to have a spin polarized ground state, we
begin by studying the emergence of magnetic ordering in
ultra-short and ultra-narrow rectangular graphene nanoribbons.  In
panels (a)-(e) of Fig.~\ref{Fig: Spin Density} we present the ground
state spin density of representative structures of each of the subsets
studied as obtained using the HSE06 density functional.  All the
structures studied have compensated lattices in the sense that the
number of carbon atoms belonging to each graphene sub-lattice are
balanced.  Therefore, according to Lieb's theorem for bipartite
lattices~\cite{Lieb1989, Rossier2007} they have no total spin moment.
The two smallest structures that present magnetic ordering in their
ground state are found to be the Phenanthro[1,10,9,8-opqra]perylene
(bisanthrene) isomer of the C$_{28}$H$_{14}$ molecule and the
Tetrabenzo[bc,ef,kl,no]coronene isomer of the C$_{36}$H$_{16}$
molecule.  The later was recently identified as spin polarized using
gradient corrected and hybrid density functionals.~\cite{Jiang2007}
This is a remarkable finding regarding the expected aromatic
hydrocarbon nature of these two isomers especially in light of a
previous study that predicted the ground state to be diamagnetic and
the first excited state to be of diradical nature (see panel (f) of
Fig.~\ref{Fig: Spin Density}).~\cite{Dias2006} Hence, our HSE06
results suggest that the energy gain from the antiferromagnetic
ordering of spins obtained in the hexaradical Clar structure (see
panel (g) of Fig.~\ref{Fig: Spin Density}) is higher than the aromatic
stabilization even for this small organic molecule.

\input{epsf}

From a qualitative analysis of the spin density maps we conclude that
the edge atoms can be divided into three subgroups: atoms which
distinctly belong to a zigzag edge, atoms that distinctly belong to an
armchair edge, and atoms belonging to the seam region between the two
types of edges (see panel (d) of Fig.\ref{Fig: Spin Density}).  
\input{epsf}
\begin{figure}[h]
\begin{center}
\epsfxsize=8.5cm \epsffile{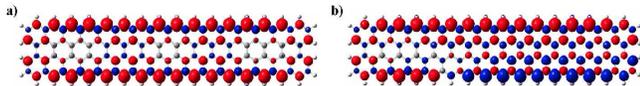}
\end{center}
\caption{Spin density isosurfaces of the $04\times 16$ GNR high spin
  multiplicity states as obtained at the PBE/6-31G** level of theory.
  Panel (a): ferromagnetic arrangement of the $m_s=3$ state. Panel (b):
  mixed configuration of the $m_s=1$ state presenting a short spin wave
  on one of the zigzag edges.}
\label{Fig: Excited States}
\end{figure}
As a rule of thumb we find that for rectangular ribbons of width $N=4$
and above, every carbon atom that distinctly belongs to a zigzag edge
will be spin polarized.  Seam region edge atoms will have a lower spin
polarization, while edge atoms on a pure armchair edge are
considerably less polarized.  Therefore, the $N\times 2$ structures,
which have no distinct zigzag edge atoms show a diamagnetic ground
state, while the $N\times 3$ (and longer) structures have magnetic
ordering in their ground state.

An interesting finding is the fact that the maximum Mulliken spin
polarization on the zigzag edge carbon atoms seems to be approximately
constant for all the systems studied and depends mostly on the density
functional chosen.  For the local density approximation this value is
usually between $0.26$ and $0.30$, whereas the maximum Mulliken spin
density is within $0.30-0.37$ at the PBE level of theory.  For the HSE
functional, the range is $0.40-0.47$.  A second important conclusion is
that for the short systems ($3 \lesssim M \lesssim 8$) the energetic
ordering of the spin states is
$E_{\uparrow\downarrow}<E_{\uparrow\uparrow}<E_0$.  Here
$E_{\uparrow\downarrow}$ is the total energy of the antiferromagnetic
ground state (Fig.~\ref{Fig: Spin Density}), $E_{\uparrow\uparrow}$ is
the energy of the ferromagnetic arrangement with a total spin moment
projection of $m_s=1$ and parallel spins on both zigzag edges (see
panel (a) of Fig.~\ref{Fig: Excited States}), and $E_0$ is the total
energy of the diamagnetic state.  For longer ribbons, while the ground
state remains antiferromagnetic, the above lying state is not
necessarily the $m_s=1$ state anymore.  The number unpaired electrons
is not sufficient to sustain the desired maximum spin polarization on
the zigzag edges while maintaining the ferromagnetic ordering.
Instead, a mixed configuration is obtained where both spin
polarizations appear on the same ribbon edge forming a type of a short
spin wave (see panel (b) of Fig.~\ref{Fig: Excited States}).  As a
result the energy of the $m_s=1$ state is raised, usually above the
$m_s=2$ state, which in turn becomes the lowest energy state with
finite total spin.

\input{epsf}
\begin{figure}[h]
\begin{center}
\epsfxsize=8.5cm \epsffile{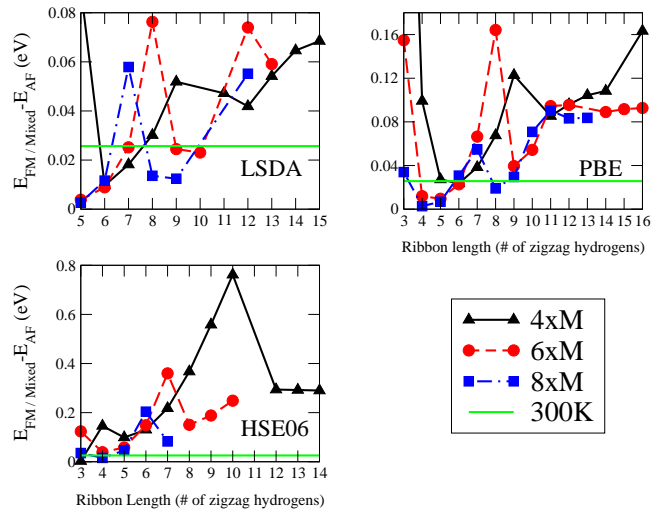}
\end{center}
\caption{Energy differences between the antiferromagnetic (AF) ground
  state and the above lying higher spin multiplicity ferromagnetic
  (FM) or mixed state for the three sets of ribbons studied as
  calculated by the local spin density approximation (upper left
  panel), PBE functional (upper right panel), and the HSE06 functional
  (lower left panel).  The green line represents $k_BT$ at room
  temperature.  Notice the different energy scales obtained for the
  different functionals calculations.}
\label{Fig: Stability}
\end{figure}

To quantify these findings we study the stability of the
antiferromagnetically ordered ground state with respect to the above
lying higher spin multiplicity state.  In Fig.~\ref{Fig: Stability}
the energy differences between the antiferromagnetic ground state and
the first higher spin multiplicity state are presented for the three
sets of ribbons studied.  All diagrams are characterized by sharp
maxima structures that correspond to the ribbon length at which a
transition in the magnetization nature of the first higher spin state
occurs, as discussed above.  We note that even though the
antiferromagnetic state of 04x03 (bisanthrene), as calculated at the
HSE06 level of theory, is only $0.003$eV ($30$K) below the above lying
ferromagnetic state, the 04x04 structure has an energy difference of
$\sim 0.15$eV ($1700$K), suggesting the detectability of its magnetic
ordering even at room temperature.  Furthermore, the HSE06 results
indicate considerably large energy differences between the
antiferromagnetic ground state and higher spin states for all ribbons
with $N\geq 4$ and length exceeding $1$nm.

\input{epsf}
\begin{figure}[h]
\begin{center}
\epsfxsize=8.5cm \epsffile{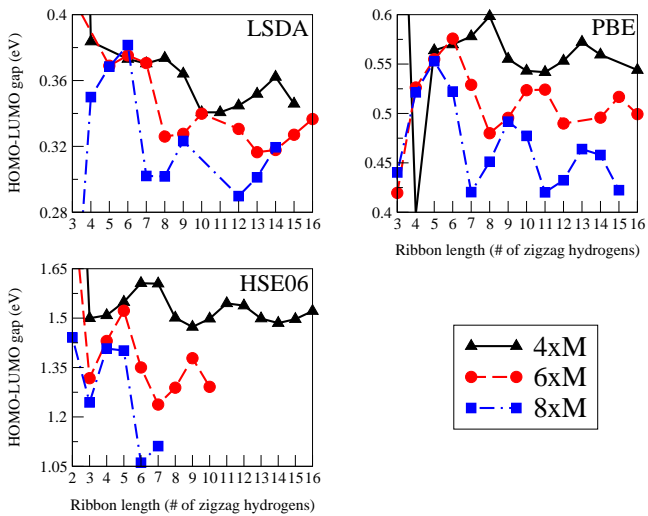}
\end{center}
\caption{HOMO-LUMO gap values for the three sets of ribbons studied as
  calculated by the local spin density approximation (upper left
  panel), PBE functional (upper right panel), and the HSE06 functional
  (lower left panel).  Notice the different energy scales obtained for
  the different functionals calculations.}
\label{Fig: Gaps}
\end{figure}
Before discussing the effect of an electric field on the electronic
properties of finite nanoribbons, it is essential to study their
ground state characteristics in the absence of external perturbations.
The geometry dependence of the HOMO-LUMO (highest occupied molecular
orbital and lowest unoccupied molecular orbital, respectively) gap
would be the most important parameter to address.  In Fig.~\ref{Fig:
  Gaps} the energy gap as a function of the length of the ribbon are
presented, for the three subsets of ribbons considered and the three
density functional approximations used.  Even though up to now we have
regarded the studied structures as zigzag nanoribbons, one can also
think of them as wide and short armchair ribbons due to their finite
length.  These type of ribbons are known to present remarkable bandgap
oscillations as a function of the ribbon width~\cite{Ezawa2006,
  Barone2006} for infinitely long armchair GNRs.  Such oscillations
can be clearly seen in Fig.~\ref{Fig: Gaps} especially for the PBE
results (upper right panel).  The periodicity of the oscillation
appears to be somewhat different than the threefold period obtained
for the infinitely long counterparts.  This, however, is a result of
the fact that in order to prevent dangling carbon bonds, the width
step taken for the finite systems is twice that taken in the
infinitely long armchair ribbon calculations.  The amplitude of the
oscillations is found to be considerably damped due to the finite size
of the ribbons in agreement with a previous study.~\cite{Shemella2007}
As expected, when the length of the armchair edge ($N$) is increased,
the oscillations amplitude increases as well.  It is interesting to
note that, in general, the HOMO-LUMO gap is inversely proportional to
the width ($N$) and the length ($M$) of the finite GNR in accordance
with the semi-metallic graphene sheet limit.  Therefore, in order to
obtain energy gap tailoring capability one will have to consider GNRs
with long armchair edges (large $N$ values) and short zigzag edges
(small $M$ values).  This will increase the amplitude of the energy
gap oscillations while maintaining overall higher gap values.

\input{epsf}
\begin{figure}[h]
\begin{center}
\epsfxsize=8.5cm \epsffile{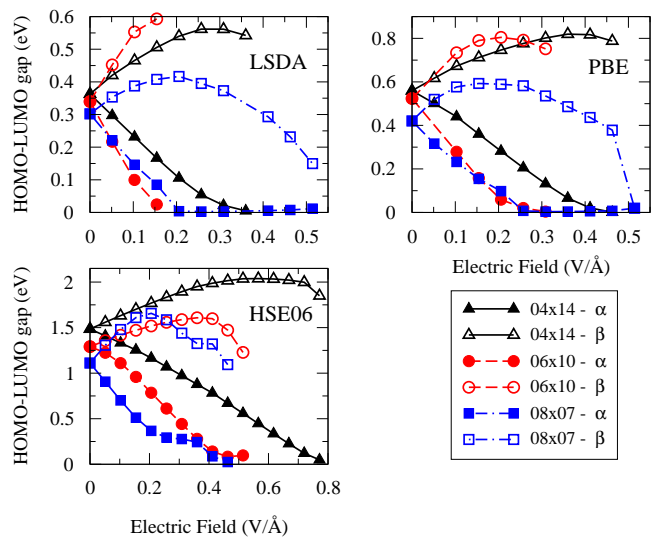}
\end{center}
\caption{Spin-polarized HOMO-LUMO gap dependence on the strength of an
  external in-plane electric field for three representative finite
  nanoribbons as calculated by the local spin density approximation
  (upper left panel), PBE functional (upper right panel), and the
  HSE06 functional (lower left panel).  Fixed geometries of the
  relaxed structures in the absence of the external field at each
  level of theory were used.  Notice the different energy scales
  obtained for the different functionals calculations.}
\label{Fig: EField}
\end{figure}

Having explored the electronic properties of unperturbed finite GNRs,
we now turn to discuss the answer to the main question addressed in
this study, namely, will finite nanoribbons turn half-metallic under
the influence of an external electric field?  In a previous study, it
was argued that the inclusion of Hartree-Fock exchange in the
approximate density functional results in the disappearance of the
half metallic state.~\cite{Rudberg2007} This was shown to be not true
in periodic systems where screened exchange (and unpublished full
exact exchange) calculations have shown half-metallic behavior similar
to that obtained with semi-local and gradient corrected
functionals.~\cite{Hod2007} In Fig.~\ref{Fig: EField} we present the
spin-polarized HOMO-LUMO gap dependence on an external in-plane
electric field applied perpendicular to the zigzag edge, for three
representative finite GNRs.  Similar to previous calculations for
periodic systems,~\cite{Son2006,Hod2007} in the absence of an external
field the $\alpha$ and $\beta$ gaps are degenerate.  Upon the
application of the field, electrons having one spin flavor experience
a smooth increase in the HOMO-LUMO gap while the opposite spin flavor
experiences a decrease in the gap.  This gap splitting continues up to
a point where the decreased gap vanishes creating a degenerate zero
energy state.  As expected, wider systems require a lower onset field
to obtain this half-metallic state.~\cite{Son2006} At this point, due
to the increased mobility of the metallic electrons, further increase
in the external field results in spin transfer between both edges thus
reducing the total spin polarization and the energy gap splitting.  At
higher electric fields the systems become diamagnetic.~\cite{Hod2007,
  Kan2007} All three representative structures present the same
features described above.  The main difference between the LSDA, PBE
and HSE06 results is the zero-field HOMO-LUMO gap, and thus the onset
electric field required to induce half-metallic behavior.
Nevertheless, they all predict the appearance of the half-metallic
state at an appropriate electric field strength.

To summarize, we have presented a detailed study of the electronic
properties of finite rectangular graphene nanoribbons.  Bisanthrene is
predicted to be the smallest organic hydrocarbon structure to present
a spin polarized ground state. This state is characterized with
antiferromagnetic ordering of spins at the to zigzag edges of the
molecule.  As a rule of thumb we find that zigzag edges, that are at
least three consecutive units long, will present spin polarization if
the width of the system is $1$nm or wider.  For systems with zigzag
edges $1$ nm and longer, our HSE06 results predict the
antiferromagnetic ordering to be considerably stable with respect to
higher spin multiplicity states, suggesting the room temperature
detectability of their spin polarization.  Depending on the dimensions
of the system higher spin states can have antiferromagnetic ordering
or be of mixed nature with spin wave characteristics.  Similar to
periodic ribbons, HOMO-LUMO gap oscillations with the length of the
zigzag edge are observed.  The amplitude of these oscillations
however, is lower than that predicted for infinite ribbons.  The
half-metallic nature of the ribbons under an external in-plane
electric field is preserved even for finite and extremely short GNRs,
regardless of the level of theory used.


\noindent
{\Large \bf  Acknowledgments}

This work was supported by NSF Award Number CHE-0457030 and the Welch
Foundation. Calculations were performed in part on the Rice Terascale
Cluster funded by NSF under Grant EIA-0216467, Intel, and HP.
O.H. would like to thank the generous financial support of the
Rothschild and Fulbright foundations.


\bibliographystyle{./jcp2.bst} \bibliography{Finite-ZZ-GNR}
\end{document}